# AN INTRUSION DETECTION MECHANISM FOR MANETS BASED ON DEEP LEARNING ARTIFICIAL NEURAL NETWORKS (ANNS)

Mohamad T Sultan[1,2], Hesham El Sayed[1,2] and Manzoor Ahmed Khan[3]

[1]College of Information Technology United Arab Emirates University, Abu Dhabi, UAE
[2]Emirates Center for Mobility Research (ECMR), United Arab Emirates University, United Arab Emirates
[3]College of Information Technology United Arab Emirates University, Abu Dhabi, UAE

*ABSTRACT*

*Mobile Ad-hoc Network (MANET) is a distributed, decentralized network of wireless portable nodes connecting directly without any fixed communication base station or centralized administration. Nodes in MANET move continuously in random directions and follow an arbitrary manner, which presents numerous challenges to these networks and make them more susceptible to different security threats. Due to this decentralized nature of their overall architecture, combined with the limitation of hardware resources, those infrastructure-less networks are more susceptible to different security attacks such as black hole attack, network partition, node selfishness, and Denial of Service (DoS) attacks. This work aims to present, investigate, and design an intrusion detection predictive technique for Mobile Ad hoc networks using deep learning artificial neural networks (ANNs). A simulation-based evaluation and a deep ANNs modelling for detecting and isolating a Denial of Service (DoS) attack are presented to improve the overall security level of Mobile ad hoc networks.*

## 1. INTRODUCTION

Recently, the significant advances in wireless networking systems have recently made them among the most innovative topics in computer technologies. Users can access a wide range of information and services through mobile wireless networks. Latest technology developments in wireless data communication devices have led to cheaper prices and larger data rates. Compared to the conventional wired networking, the wireless networking provides a great deal of flexibility, efficiency and cost effectiveness that make them a good alternative in providing an efficient network connectivity. The development of Mobile Ad hoc networks (MANETs) [1][2] presented a reliable, cost-effective and efficient techniques exploit the availability and presence of mobile hosts during the lack of a fixed communication infrastructure. In MANET, the mobile nodes are independent and can effortlessly initiate a direct communication channel with each other as they are freely moving around the infrastructure-less network in different directions and at different velocity speeds. The Ad hoc network functions in a very specific way and nodes cooperation is its main element for forwarding the communication related information from main data sources to the planned destination mobile nodes. Nodes in MANET relies entirely in its operation on batteries as means of energy to move arbitrarily with no restrictions. The mobile nodes could leave or join the dynamic network at any specific time and can take independent decisions without relying on any centralized authority. Due to its core aspect of abandoning the availability of any fixed infrastructure as a necessary factor for the communication to be present. This has dictated that the transmission and communication range of the entire network will be determined by the transmission range of the individual mobile nodes, and it is usually smaller in size compared to the range of the cellular networks. Nevertheless, in cellular networks to avoid interference and provide guaranteed bandwidth, each communication cell depends on various communication frequencies available from the on-hope adjacent neighbouring cells. This expands the communication range of the cellular network especially when different communication cells are joined together to offer a radio and communication coverage for wide-ranging geographical area. However, in MANET each mobile node has a wireless interface and interconnects with other nodes over a wireless channel. Mobile nodes in MANET could range from portable laptops to smartphones or any other digital devices with a communication wireless antenna. Among the numerous advantages the infrastructure-less ad hoc network offers are robustness, efficiency, and inherent support for dynamic random mobility. Fig. 1 illustrates the architecture of MANET.

The special characteristics of MANETs have made its deployment a preferable choice for many fields such as in military battlefield operations, natural disasters and in remote areas. However, due to their openness and decentralized structure. MANET has become vulnerable to different kinds of malicious threats and attacks. Their flexibility brings new threats to its security. The categorization of threats that affect mobile ad hoc networks can be perceived in many ways based on behaviour, level and position of the specific attack, flows of the used security algorithms and weaknesses in the structure of the developed routing protocols. Attacks such as blackhole attack, network partition, node selfishness, malicious node, and denial of service (DoS) are among the many popular threats that MANETs is facing [3]. The shared goal for those threats is to degrade the overall network performance. Researchers have become more focus on how to enhance and provide a secure and reliable mobile ad hoc network. Several techniques have been developed such as signature-based, statistical anomaly-based, and protocol analysis. This research will focus on deep learning intrusion detection techniques in MANET based on artificial neural networks The paper concentrates on very specific attack which is the denial-of-service DOS attach that can easily disrupt the MANETs operations.

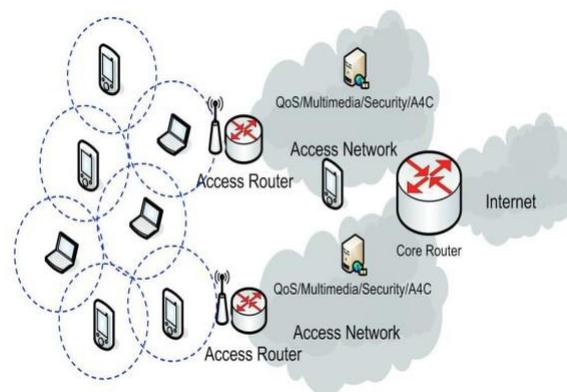

Figure 1. MANET architecture

## 2. RELATED WORK

Identifying malicious and misbehaving mobile nodes is necessary to protect the MANET network. Researchers have conducted research on studying the security threats of mobile ad hoc network to make MANET more secure and reliable. In describing the security threats, many researchers make their own categorization of the security threats. MANET threats are classified into two levels. The first level is attacks on the basic mechanism resulted from nodes captured, compromised or the misbehaviour of nodes that do not listen to the rules of cooperative algorithms. The second level is attacks on the security mechanism which exploit the vulnerabilities of the security mechanism employed in MANET. In [4] and [5] researchers have classified security attacks in correspondence with the communication layers, which mean that each layer has its own threats and vulnerabilities. Table 1 shows security threats at the communication layers.

Table 1. Communication Layers Security Threats

| Layers | Attacks |
| --- | --- |
| Application layer | Selfish nodes attacks, Malicious attacks like viruses, worms and spyware. |
| Transport layer | Session hijacking, Session control, flooding attack and ACK-storm attack on TCP |
| Network layer | Cache poisoning attacks, Routing protocols attacks (e.g., AODV, DSR, TORA), Packet dropping attacks, blackhole attack node impersonation, denial-of-service DoS attack. |
| Data link layer | Man in the middle attack, MAC interruption (802.11), WEP vulnerability. |
| Physical layer | Eavesdropping, jamming, traffic interceptions. |

In [6] the authors have studied the effect of misbehaviour nodes on the MANET network. In this research a new method was used to efficiently detect and separate malicious mobile nodes from the network. Thus, the network performance remains balanced and stable regardless of the presence of the colluding nodes. The malicious behaviour is represented by suspicious behaviour of unauthorized mobile nodes that can inflict damage on other nodes in the network intentionally or unintentionally. An example of this could include the scenario where the aim of the mobile node is not the attack itself but to gain unauthorized benefits over other nodes.

The researchers in [7] proposed a blackhole attack identification mechanism in MANET using fuzzy-based intrusion detection techniques. Their main target was to detect the blackhole attack in the mobile network, which is considered as very popular type of malicious attack that disrupts the operations of MANETs. An adaptive neuro fuzzy inference system was developed by the researchers. The development of this system was based on the popular optimization technique; particle swarm optimization (PSO). Similarly, using fuzzy logic techniques the authors in [8] used a new technique for intrusion detection called node blocking mechanism, to differentiate two popular attacks that targets the network which are the grey hole and the black hole attacks,
The authors in [9] proposed a system that uses malicious behaviour-detection ratios to enhance security in mobile networks using modified zone-based intrusion detection techniques. In [10] another intrusion detection system was proposed using smart approach for intrusion identification and isolation. This system detects an attack on the ad hoc network by exhibiting a deep learning neural network with bootstrapped optimistic algorithm. In this system each mobile node submits finger vein biometric, user id, and latitude and longitude then the intrusion detection is executed to verify these entities and detect any suspicious behaviour in the network.

## 3. MANET ROUTING PROTOCOLS

The random arbitrary nature of mobile nodes in MANET due to the absence of any fixed communication infrastructure keep the network's topology in constant change. This rapid and dynamic change in topology make routing in MANET a challenging task. Thus, an effective routing strategy is required to smoothly accomplish the forwarding of packets form the source to the destination. The routing information is changing frequently to reflect the dynamic changes in network topology. There are numerous potential paths from source to destination. The routing protocol algorithms discovers a route and transports the data packets to the appropriate destination. Numerous routing algorithms for MANET have been developed [11]. The performance of the ad hoc network is highly associated with used routing protocols efficiency. The proposed routing algorithms for MANET can be divided into three different categories based on their behaviour and functionality. These categories are proactive, reactive and hybrid routing algorithms [11][12]. The basic concept of these routing algorithms is to discover the shortest route for the source-destination routes selection.

### 3.1. AODV and Targeted Attacks

In MANETs, one of the widely used routing protocols that follows the reactive routing mechanism is the Ad hoc on demand distance vector (AODV) [13]. The AODV protocol sets up routes using a query cycle consisting of Route request (RREQ) and Route Reply (RREP). If a node has the most recent sequence number for a certain destination and needs to deliver data packets to that location, it will broadcast an RREQ message to its neighbours. Until the requested data is available in some form, this message will be transmitted. After receiving the RREQ message, every node builds a path back to its original sender. After receiving an RREQ message, the destination will respond with an RREP message that includes the destination's current sequence number and the number of hops taken to get there [14][17]. Keep in mind that if a given intermediate node has a newly discovered route to the final destination, it will not relay the RREQ to its neighbours but will instead send an RREP back in the direction of the source. Each node that gets the RREP message sets up a new forward route to the final destination. Thus, rather of storing the whole path, each node simply stores the information necessary for the next step. When a node detects that it has received a duplicate RREQ, it discards the packet. As an added measure, AODV verifies the freshness of the routes by using sequence numbers. The destination routes are only altered if a new path to a given destination has a higher sequence number than the previous path or has the same sequence number but with fewer hops. Moreover, when there is a link failure or a routing problem happens in the network, another technique is executed in AODV which is the route error (RERR) [13][14]. This technique sends warning error packets to the source and destination nodes in the ad hoc network. As an example, used in this research, this section discusses examples of attacks on AODV routing protocol.

- Packet dropping attack: In this type of attack malicious mobile users may drop all the legitimate incoming data packets that are mainly employed in route discovery and route maintenance stages. This is usually happening with aim of disrupting the network services such as (RREQ, RREP and RERR).
- Denial-of-Service (DoS): One of the most popular known attacks. This type of network attack, render a resource or a service inaccessible in the network [15]. The main aim of this malicious attack is not to get an unauthorized access by the perpetrator but is rather an act of vandalism to shut down a machine, resource or network. This attack will usually result in a legitimate users being unable to access the available resources. In AODV routing algorithm the attacking malicious node that wants to disrupt MANET resources begins to frequently broadcast the route request (RREQ) messages while the route discovery process is taking a place. Using an expired destination IP address that is still available in the address range of the network but is not present in the mobile network anymore the neighboring nodes receive these malicious messages. This malicious activity results in deceiving the mobile nodes to re-forward the route request messages as there is no neighboring node having a fresh route to this fake and expired destination address. The main goal of DoS attack is to consume the battery power of nodes and disrupt or deny access of legitimate nodes to a specific network services.

## 4. METHODOLOGY

The mobile ad hoc networks are highly vulnerable to different kinds of security threats due to its dynamic nature of randomness, decentralizations, and lake of central authority. The aim of this work is to propose an effective intrusion detection mechanism for MANETs using deep learning techniques such as artificial neural networks (ANNs). The denial-of-service attack that is being considered in this research is implemented in way where a malicious intruder mobile node injects its malicious data packets in large volumes into the mobile ad hoc network which leads to a disruption and denial of services at the destination node. The main routing protocol used in this study to perform the simulation experiments is the AODV. This routing protocol is used due to its popularity in MANET. In this study and in our experimental setup all the factors and issues that has an impact on link stability on the network is considered and analysed. The main attack which is considered in this research is the Denial of Service (DoS) attack with the aim of rendering the MANET resources and services inaccessible by overloading it with junk packets in a two way network communication setting. This type of attack has the possibility to happen over both wired and wireless networks. However, the wireless networks are more susceptible due to its radio nature and more loosely specified restrictions such as the case of MANETs. Fig. 2 shows an example of DoS attack where intruder D floods the host node C with extra malicious packets.

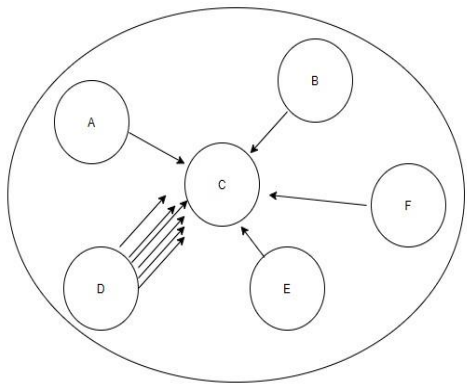

Figure 2. Example of DoS attack, the host node C flooded by Intruder D

The deep learning ANNs are used to detect intrusions based on abnormal network activity and the attributes, labels and features are selected from the packets generated during the network simulation. Given to the learning and generalizable attributes of artificial neural networks (ANNs), and due to their ability to obtain knowledge from data and infer new information, are more suitable to manage such tasks. The performance of the proposed intrusion detection system is illustrated by means of simulation using AODV routing protocol in MANETs. ANN modelling for attack detection using a simulated MANET environment will be used in this research.

## 4.1. Implementation

The implementation of the proposed research is illustrated by means of simulations using NS-2 network simulator on Linux ubuntu 10.04 platform to evaluate the performance of the MANET network with 15 mobile nodes forming a network. Attack detection, using a simulated MANET environment and ANNs modelling is used. Once a simulation process is completed, NS2 follows to display the simulation details is by generating a big sized trace file holding all the events sequentially line by line. For all those reasons, the event-driven technique is used in NS2 as it can keep all the occurred events as records and all those records can be traced and analysed for evaluation purposes. In NS2 there are typically two kinds of output data records that can help in further investigation for a specific simulation scenario. The first one is a trace file which records the events traces that assist in studying the performance of the network by processing and analysing it using numerous of methods. The second one is a network animator (NAM) file which assists in detecting the interactions and movements between the mobile nodes visually. Fig. 3 illustrates the complete procedure of how a specific simulation is conducted using NS2.

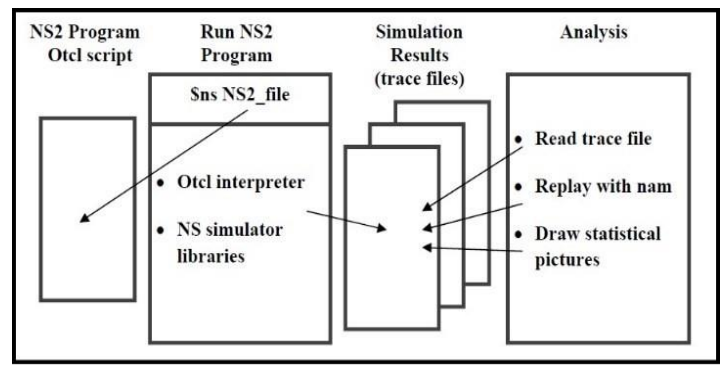

Figure 3. NS2 simulation process

### 4.1.1. Mobility Model

The mobility model plays a very important role in MANET simulations. The considered model should attempt to simulate the movement, behaviour, and actions of real nodes in MANET. However, the mobile nodes in MANET move in a very dynamic arbitrary and decentralized manner. It's a dynamic network of autonomous decentralized mobile nodes. A node in the network could join or leave at any specific time which leads to high rates of link and topology changes. Moreover, the mobile nodes make decision independently and behave as routers where they can send, receive, or route the information simultaneously. Thus, to model this kind of unpredictability and randomness that exist in mobile ad hoc networks, the researchers have proposed different probability distribution models of MANET nodes. The most popular one that is highly represent the distribution of MANET nodes is called the Random Waypoint Mobility Model [16]. For this model the spatial distribution of mobile nodes movements is in general a non uniform. The mobility model that represents the movements of the mobile nodes is an important aspect for any simulation process because the way that these mobile nodes move and behave affects in different ways the performance of the routing protocol that these nodes utilize. The random waypoint mobility model is simple, reliable and is highly used to assess the behaviour of the MANET [16]. This mobility model can highly represent the actions and movements of real mobile nodes in real conditions. There is basically a specific pause time in this model that operates when there are any sudden changes or differences in direction or velocity of mobile nodes. When a specific wireless node starts to travel across the network, it remains in one location for a particular period of time which is a pause time before it moves to another location.

The node chooses the subsequent destination randomly in the simulation region once that specified pause time has expired. These mobile nodes also select a speed that is generally specified between the minimum and maximum speed (0, Maxspeed) during simulation process. Then it travels to the newly chosen point at that selected speed. When the mobile node reaches at targeted place, it starts waiting again for a certain period of time, seconds before selecting another new way point and another speed. Then it initiates the same procedure all over again. Numerous researchers have adopted and implemented this mobility model in their studies. The Movement of individuals in a cafeteria or shopping mall, and movement of nodes in a conference are some of its practical examples. Fig. 4 presents an illustration of the movement pattern for a mobile node which begins at a randomly selected location (133, 180) and chosen a speed between (0 to 10 m/s) using random waypoint mobility model.

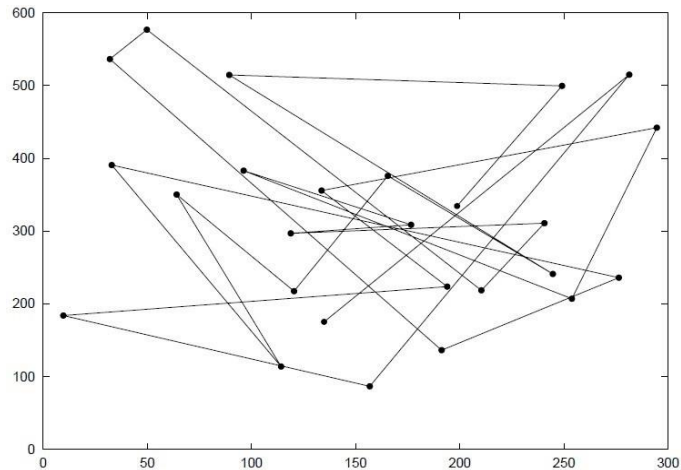

Figure 4. Random waypoint mobility for node movement pattern

## 4.2. Simulation Setup and Parameter Selection

A scenario file that defines the exact motion of every node in the network along with the exact number of packets generated by each node in the network is being taken as an input for every simulation run. This is together accompanied by the exact time at which each change in motion or packet origination is to occur. The simulation is done using NS2 simulator as shown in table 2 below:

Table 2. Simulation Parameters

| Simulation parameters | |
|---|---|
| **Parameter** | **Selected Value** |
| Routing Protocol | Ad hoc on demand distance vector (AODV) |
| Platform | Linux distribution ubuntu version 10.04 |
| Number of Nodes | 15 |
| Simulation Software | NS-2 |
| MAC Layer Protocol | IEEE 802.11b |
| Simulation Area | 500m X 500m |
| Traffic Generation Model | CBR (Constant Bit Rate) |
| Size of packet | 512 bytes |
| Mobility Model | Random Waypoint |
| Maximum Speed | 0-20 m/s |
| No. of Connections | 2 to 10 |
| Duration of experiment | 200 sec |
| Type of Antenna | Antenna/OmniAntenna |

In this simulation execution, 15 nodes are deployed for MANET within the terrain of 500m X 500m using random waypoint mobility for the purpose of realization of a real-time simulation and the simulation runs for the maximum experiment duration of 200s, with maximum speed of 20m/s. It's indicated in the simulation parameters table the "Maximum Speed" of mobile nodes which is in fact implies that the node's speed is already changing form "0 m/s" which is a stationary paused node "no movement" to maximum speed of "20 m/s". Since we have used the popular mobility model "Random Waypoint Mobility Model", which is designed to specify users or mobile nodes movement, their location, velocity, and acceleration change over time. The mobile nodes speed in our simulation environment could change at any random time form (0 – 20 m/s). The MAC layer used is IEEE 802.11b. Once the simulation is finished, the generated output files like trace files should be analysed to extract beneficial data and statistics. As stated earlier, a pair of files will be produced once the simulation process ends. The first one is an event trace file which records all simulation events while the second one is a network visualization file which records the data that can be used in network animation. These event trace files are in its raw format and an analysis and assessment should be performed in order to extract the required necessary information. Both files are CPU intensive tasks while in simulation process and they make use and occupy an amount of the memory. The example excerpt in Fig. 5 below shows how a generated trace file will look like after a simulation run.

```
s -t 2.556838879 -Hs 1 -Hd -2 -Ni 1 -Nx
342.47 -Ny 4.35 -Nz 0.00 -Ne -1.000000 –
Nl RTR -Nw --- -Ma 0 -Md 0 -Ms 0 -Mt 0
```

Figure 5. Excerpt of trace file

The excerpt above indicates that the data packet was sent (s) at time (t) 2.556838879 sec, from the main source node (Hs) 1 to target mobile node (Hd) 2. The source node id (Ni) is 1, The source node X axis coordinates (Nx) is 342.47, while the provided Y axis coordinate (Ny) is 4.35.

Moreover, its Z coordinate (Nz) is 0.00. The available level of energy (Ne) is 1.000000, while the type of trace format for this mobile node for routing (Nl) is RTR and the event of the node (Nw) is blank. Moreover, (Ma) 0, is the specification of MAC level information while the address of the destination Ethernet (Md) 0, the address of the source Ethernet (Ms) is 0 and Ethernet kind (Mt) is 0. The features extracted from the logged details can then be used in ANN for attack detection. The analysis process of these trace files can be done using different tools such as using the AWK language command and Perl scripts. Different parameter selection for data extraction can be considered for analysis which merely depends on the nature of the network and the specific attack. The following parameters will be considered: Packet Loss PL, Packet sent (PS), Packet received (PR), Energy consumption (EC). Using analysis log files of simulation run, the parameters were extracted. The data is split for training and testing where 65% of data including 15 mobile nodes in 200 seconds were selected randomly for training and 35% for the purpose of testing and validation process.

### 4.3. Designing Artificial Neural Network

An intrusion detection system using neural network (NN) is proposed to secure the MANET. Neural Network model is trained by applying the simulation data as inputs to the ANN. Feed Forward Back Propagation (FFBP) in the Neural network toolbox is used and the artificial neural network is implemented with four inputs, one output layer including two middle hidden layers. The network training in this setup is conducted using back propagation (BP) learning process, The (TRAINLM) training function of Levenberg-Marquardt backpropagation is used in addition to LEARNGDM as an adaptive learning function. Different transfer functions are available like Purelin, Log-Sigmoid, and Tan-Sigmoid. The main aim of the transfer function is to be used for estimating the output of a specific network layer from its initial net input. LogSigmoid, and Tan-Sigmoid are used in this study. Fig. 6 below shows an example of different transfer functions.

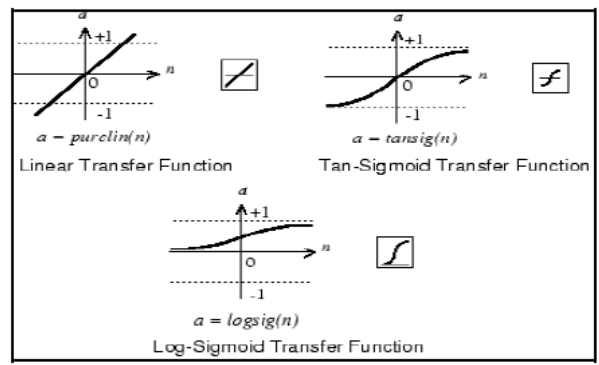

Figure 6. Example of different transfer functions

A screenshot of how the artificial neural network setup and design is presented in Fig 7 and Fig. 8 respectively. All the setup parameters must be specified before running the artificial neural network.

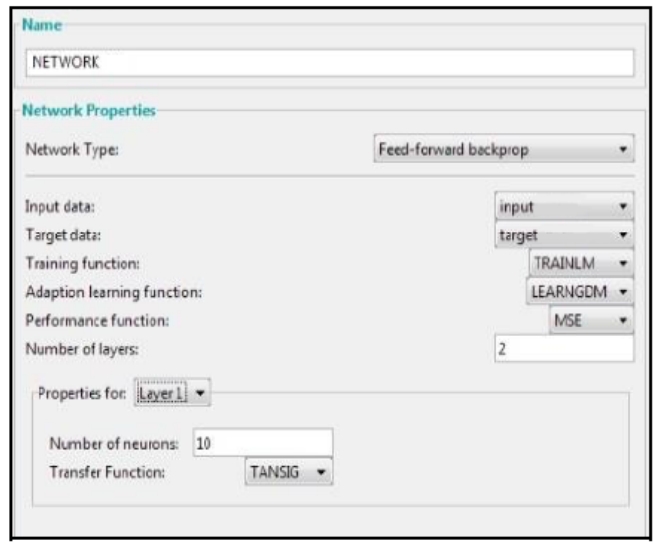

Figure 7. Neural network setup

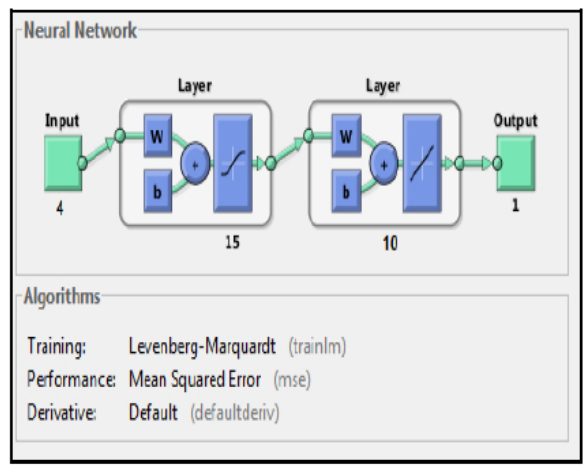

Figure 8. Neural network design

## 4.4. Modelling Artificial Neural Networks for DOS Attack Detection

Given to the learning and generalizable attributes of feedforward neural networks with back propagation training algorithm, those deep learning networks are used for the purpose of DoS intrusion detection and to identify and predict any unusual activity and the features are selected from the packets generated in the simulation process. The number of input nodes will be determined from the input data set. The number of nodes in the hidden layers in the neural network are varied frequently during the experiments to achieve a highly accurate and stable neural network model and to avoid any overfitting. The structural design of the proposed deep learning neural network consists of two types of different network setups. The first one has 4 inputs and 15 neurons in the first hidden layer and 10 neurons in the second hidden layer and one output. While the second network has 4 inputs and 20 neurons in the first hidden layer and 10 neurons in the second hidden layer and one output. Training using feed forward back propagation (FFBP) in ANN is presented in Fig. 9 and the process is indicated as follows:

- The model selects training epoch from the training set and initialize weights and biases.
- The model the calculates the output of the network.
- Then, the error between the network output and the desired output is calculated.
- The model modifies the weights of the network in a way that minimizes the error.
- The model repeats the steps for each input in the training set until the error for the entire set is acceptable low.

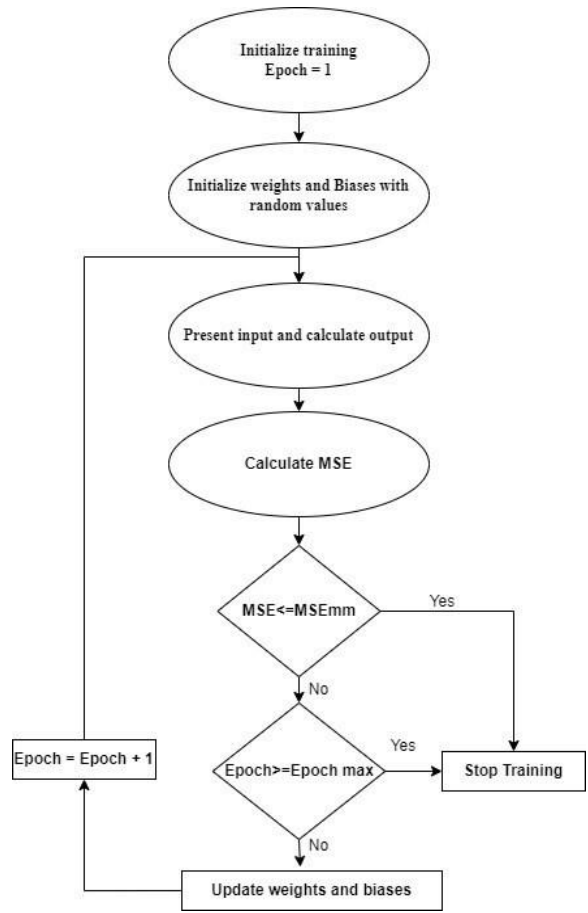

Figure 9. Training process

## 5. PERFORMANCE RESULTS

The deep learning technique which is used to design the ANN uses the backpropagation training algorithm to predicts a specific output. Then, this output is compared with actual known class label to measure the difference in error between the predicted and actual outputs. The obtained error is sent back to the neurons for adjustments. FFBP measures the variance of the residuals in a repeated process. The root mean squared error is just one way to calculate this error. The method of squaring the sum of the error is used to prevent the cancel out the positives and negatives values during the sum of the error of all the nodes. We used the root mean squared error instead of the mean absolute error (MAE) to measure the standard deviation of errors as the gradient descent requires the derivative of that loss function to be calculated to minimize the loss function and generate better outputs. The results are presented in table 3 below. The performance results of the designed deep learning model are shown in the table based on the training data. The selection process of the best performing model is based on results obtained.

Table 3. Artificial neural network for training data

| Network | Training/ Learning functions | Layers | Transfer function | RMSE | Epoch |
|---|---|---|---|---|---|
| Feed Forward Back Propagation (FFBP) | Training: TrainLM  Learning: LearnGDM | 4–15–10-1 | LogSigmoid | 0.1924 | 10 |
| | | | LogSigmoid | 0.1901 | 12 |
| | | | TanSigmoid | 0.0452 | 14 |
| | | | TanSigmoid | 0.0618 | 16 |
| | | 4-20-10-1 | LogSigmoid | 0.1927 | 10 |
| | | | LogSigmoid | 0.1801 | 12 |
| | | | TanSigmoid | 0.0492 | 14 |
| | | | TanSigmoid | 0.0835 | 16 |

We executed the neural network to detect unusual malicious activities in MANET. As it can be noticed in the performance results that two different transfer functions are used in this research Log-Sigmoid and Tan-Sigmoid. A well-trained ANN should have a very low RMSE at the end of the training phase The best result in ANNs for FFBP network with Tan-Sigmoid function is related to 4-15-10-1 network that produce RMSE=0.0452, for 14 epochs. The indication of MSE being quite small or almost close to zero is that the neural network model output and the desired output have become very close to each other for the training dataset. The rest of results are given in table 4 below. The table shows the performance results of the neural network model based on testing data. It can be noticed that the best result for neural network model using FFBP with Tan-Sigmoid function is related to 4-15-10-1 design that produce RMSE=0.0512.

Table 4. Artificial neural networks based on testing data

| Network | Training Function | Layers | Transfer function | RMSE |
|---|---|---|---|---|
| Feed Forward Back Propagation (FFBP) | Training: TrainLM  Learning: LearnGDM | 4–15–10-1 | LogSigmoid | 0.1998 |
| | | | LogSigmoid | 0.1982 |
| | | | TanSigmoid | 0.0512 |
| | | | TanSigmoid | 0.0781 |
| | | 4-20-10-1 | LogSigmoid | 0.2337 |
| | | | LogSigmoid | 0.1891 |
| | | | TanSigmoid | 0.0821 |
| | | | TanSigmoid | 0.0935 |

Both of Fig. 10 and Fig. 11 show a summary of how the designed artificial neural networks (ANN 4-15-10-1) and (ANN 4-20-10-1) performed for training and testing phases. In this research, after we selected the best model with best RMSE value, we used this model to evaluate the performance of proposed system. The goal is to distinguish a normal connection form a malicious attack connection in MANET. Thus, we used a performance measure which is the Detection Rate (DR). This measure is calculated as the number of attack connections which classified correctly as an attack over the total number of connections in the network. Using this measure, we were able to detect the attack in the network with high accuracy as shown in the table below. It can be noticed that as the number of connections increases the detection rate decreases due to higher false positive rates.

International Journal of Computer Networks & Communications (IJCNC) Vol.15, No.1, January 2023

Table 5. Detection Rate

| No. of connections | Detection Rate |
|---|---|
| 2 | 95.3% |
| 5 | 94.2% |
| 10 | 91.73% |

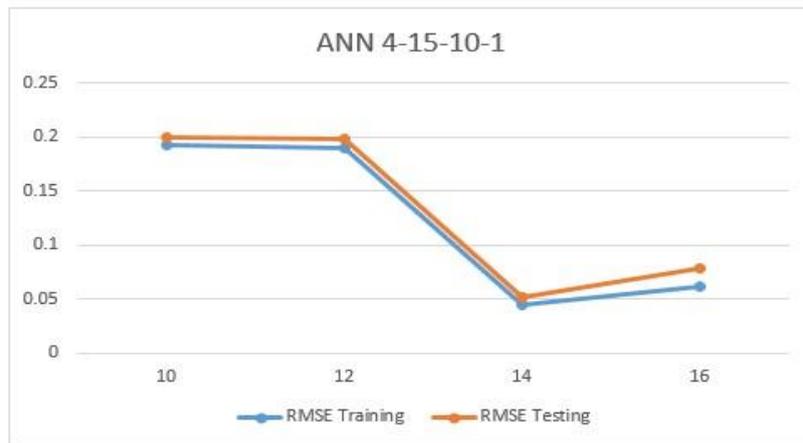

Figure 10. RMSE for Training and Testing (ANN 4-15-10-1)

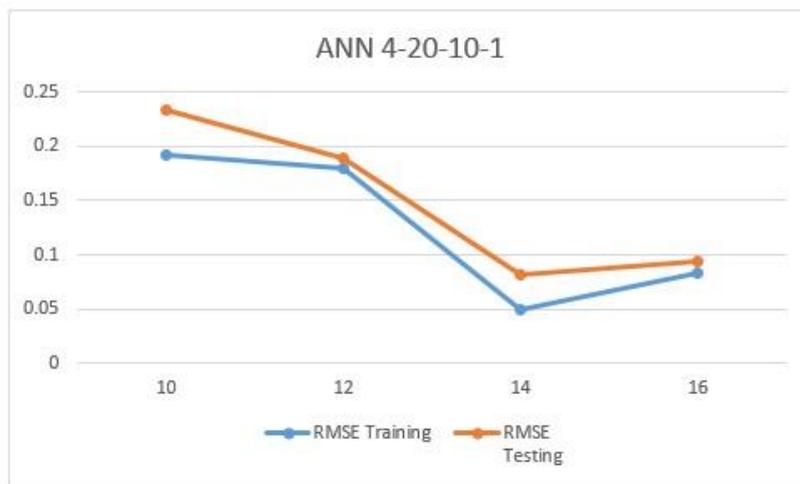

Figure 11. RMSE for Training and Testing (ANN 4-20-10-1)

## 6. CONCLUSIONS

This research paper is mainly focused on modelling and investigating the use of artificial neural networks ANNs as a mean for intrusion detection in mobile ad hoc networks (MANETs). The main objective of this work was to analyse, simulate and evaluate the use of feedforward neural networks with back propagation (FFBP) in MANETs. An extracted dataset generated using the means of simulations for mobile ad hoc networks is used to calculate the input parameters of this approach and the RMSR is employed as metric to evaluate the performance of the proposed deep learning artificial neural network modelling. The proposed modelling can be utilized for detecting

DoS attack in MANET. The best results in ANNs for FFBP network with Tan- Sigmoid function is related to 4-15-10-1 network that produce RMSE=0.0452, for 14 epochs for training and RMSE=0.0512 for testing data. We also used the Detection Rate (DR) as a performance measure to evaluate the selected neural network model. For the future works, different types of network attacks will be considered for the purpose of intrusion detection. Another measure could be used in the analysis is the coefficient of determination or R squared. R square is the percentage of variation in Y explained by the model. The higher the percentage of the R square is the better. However, the value of R square will be always less than one irrespective of the values in dataset being small or large.

## CONFLICTS OF INTEREST

The authors declare no conflict of interest.

## ACKNOWLEDGEMENTS

This research was funded by the Emirates Center for Mobility Research (ECMR) of the United Arab Emirates University (grant number 31R271).

## REFERENCES

[1] Baltaci, A., Dinc, E., Ozger, M., Alabbasi, A., Cavdar, C., & Schupke, D. (2021). A Survey of Wireless Networks for Future Aerial COMmunications (FACOM). IEEE Communications Surveys & Tutorials.
[2] Srilakshmi, U., Veeraiah, N., Alotaibi, Y., Alghamdi, S. A., Khalaf, O. I., & Subbayamma, B. V. (2021). An improved hybrid secure multipath routing protocol for MANET. IEEE Access, 9, 163043-163053.
[3] Meddeb, R., Triki, B., Jemili, F., & Korbaa, O. (2017, May). A survey of attacks in mobile ad hoc networks. In 2017 International Conference on Engineering & MIS (ICEMIS) (pp. 1-7). IEEE.
[4] Korba, A. A., Nafaa, M., & Salim, G. (2013, April). Survey of routing attacks and countermeasures in mobile ad hoc networks. In 2013 UKSim 15th International Conference on Computer Modelling and Simulation (pp. 693-698). IEEE.
[5] Yang, H., Luo, H., Ye, F., Lu, S., & Zhang, L. (2004). Security in mobile ad hoc networks: challenges and solutions. IEEE wireless communications, 11(1), 38-47.
[6] Pragya, M., Arya, K. V., & Pal, S. H. (2018). Intrusion detection system against colluding misbehavior in manets. Wireless Personal Communications, 100(2), 491-503.
[7] Moudni, H., Er-rouidi, M., Mouncif, H., & El Hadadi, B. (2019). Black hole attack detection using fuzzy based intrusion detection systems in MANET. Procedia Computer Science, 151, 1176-1181.
[8] Balan, E. V., Priyan, M. K., Gokulnath, C., & Devi, G. U. (2015). Fuzzy based intrusion detection systems in MANET. Procedia Computer Science, 50, 109-114.
[9] Krishnan, R. S., Julie, E. G., Robinson, Y. H., Kumar, R., Son, L. H., Tuan, T. A., & Long, H. V. (2020). Modified zone based intrusion detection system for security enhancement in mobile ad hoc networks. Wireless Networks, 26(2), 1275-1289.
[10] Islabudeen, M., & Devi, M. K. (2020). A smart approach for intrusion detection and prevention system in mobile ad hoc networks against security attacks. Wireless Personal Communications, 112(1), 193-224.
[11] Chen, Z., Zhou, W., Wu, S., & Cheng, L. (2020). An adaptive on-demand multipath routing protocol with QoS support for high-speed MANET. IEEE Access, 8, 44760-44773.
[12] Tahboush, M., & Agoyi, M. (2021). A hybrid wormhole attack detection in mobile ad-hoc network (MANET). IEEE Access, 9, 11872-11883.
[13] Saini, T. K., & Sharma, S. C. (2020). Recent advancements, review analysis, and extensions of the AODV with the illustration of the applied concept. Ad Hoc Networks, 103, 102148.
[14] Darabkh, K. A., Judeh, M. S., Salameh, H. B., & Althunibat, S. (2018). Mobility aware and dual phase AODV protocol with adaptive hello messages over vehicular ad hoc networks. AEU International Journal of Electronics and Communications, 94, 277-292.

[15] Wu, C., Wu, L., Liu, J., & Jiang, Z. P. (2019). Active defense-based resilient sliding mode control under denial-of-service attacks. IEEE Transactions on Information Forensics and Security, 15, 237-249.
[16] Singh, A., Sharma, S., & Srivastava, R. K. (2020). Investigation of random waypoint and steady state random waypoint mobility models in NS-3 using AODV. Journal of High Speed Networks, (Preprint),1-8.
[17] Chandan, R. R. (2020). Consensus routing and environmental discrete trust based secure AODV in MANETs. International Journal of Computer Networks & Communications (IJCNC) Vol, 12.